\documentclass[aps,twocolumn,nofootinbib]{revtex4}

\usepackage{natbib}

\usepackage{graphicx}
\usepackage{amssymb}
\usepackage{subfigure}
\usepackage{stackrel}
\usepackage{amsfonts}
\usepackage{bm}
\usepackage{amsmath}
\usepackage{xcolor}
\usepackage[normalem]{ulem}
\usepackage{hyperref}
\usepackage{verbatim}
\definecolor{rosita}{rgb}{0.97, 0.56, 0.65}
\definecolor{verde}{rgb}{0, 0.56, 0.65}

\newcommand{\ket}[1]{\ensuremath{\left|{#1}\right\rangle}}
\newcommand{\bra}[1]
{\ensuremath{\left\langle{#1}\right |}}

\newcommand{\sdo}{\ensuremath{\downarrow}}
\newcommand{\su}{\ensuremath{\uparrow}}

\newcommand{\be}{\begin{equation}}
\newcommand{\ee}{\end{equation}}
\newcommand{\ben}{\begin{eqnarray}}
\newcommand{\een}{\end{eqnarray}}

\newcommand{\ana}[1]{{\color{magenta} #1}}

\begin{document}
\title{Generation of tripartite entangled states with fermionic systems}

\author{M. D. Jim\'{e}nez$^1$} 
\author{A. Vald\'{e}s-Hern\'{a}ndez$^2$}\email{andreavh@fisica.unam.mx}
\author{A. P. Majtey$^1$ }

\address{$^1$Instituto de F\'{i}sica Enrique Gaviola, CONICET and Universidad 
Nacional de C\'{o}rdoba,
Ciudad Universitaria, X5016LAE, C\'{o}rdoba, Argentina.\\$^2$ Instituto de F\'{\i}sica, Universidad Nacional Aut\'{o}noma de M\'{e}xico, Apartado Postal 20-364, Ciudad de M\'{e}xico, Mexico.}

\begin{abstract}

 We propose a protocol based on a tunneling plus particle-detection process aimed at  generating tripartite entanglement in a system of 3 indistinguishable fermions in a triple-well potential, initially prepared in a state exhibiting only exchange correlations. 
Particular attention is paid to the generation of \emph{fermionic} \textsc{ghz}- and \textsc{w}-type states, which are analogous to the usual \textsc{ghz}- and \textsc{w}-type states defined in composites of distinguishable qubits.  
The protocol succeeds in generating fermionic \textsc{w}-type states,  
and the ensuing state becomes effectively equivalent to a 3-distinguishable-qubit \textsc{w}-type state shared among three localized parties.  
The protocol, however, is unable to generate \textsc{ghz}-type states, a result that highlights the fundamental inequivalence between these two types of states, and throws light into the characterization of processes that guarantee the emergence of specific kinds of  multipartite entanglement in systems of identical parties.
Our findings suggest new paths for the exploration, generation and exploitation of multipartite entanglement in composites of indistinguishable particles, as a useful resource for quantum information processing.

\vspace{5mm}

\end{abstract}

\maketitle

\section{Introduction}

The creation and manipulation of entangled multipartite states play a pivotal role in quantum information processing, and provide a powerful resource for various quantum technologies \cite{nielsen_chuang_book,Jozsa2003,horodecki_2009}. 
Multipartite entanglement also acquires importance in the study of fundamental questions in quantum mechanics, specifically related to the nature of quantum correlations. 
One particularly active line of research focuses on the generation of entanglement in systems of indistinguishable particles, out of the exchange correlations due to the symmetry of the composite state. 

When studying quantum systems composed of many identical particles, different ways to characterize entanglement arise.
One possible approach focuses on mode entanglement, by partitioning the system into distinguishable occupation states \cite{zanardi_2000,Benatti_2020,Dalton_Identical_Bosons_I}. 
An alternative avenue considers entanglement between the particles themselves, which becomes useful mode entanglement when distributed to separate parties \cite{Morris_2020}. 
Such alternative, however, requires revisiting the definition of separability in the case of indistinguishable fermions, mainly because the state's exchange antisymmetry prevents pure states of identical fermions from being factorizable into products of single-particle state vectors.  
Here we adopt the notion of separability  according to which a generic quantum state is regarded as separable, or non-entangled, whenever a complete set of physical properties  can be attributed to all individual subsystems \cite{Ghirardi_2002,Majtey2023}. 
For indistinguishable-fermion systems, this amounts to define \emph{ fermionic entanglement} as those  correlations \emph{on  top} of the exchange correlations, that is, as correlations beyond those contained in a Slater determinant \cite{schielmann_2001,Ghirardi_2002,Eckert_2002,plastino_2009_epl}.

The role of  exchange correlations as a resource for quantum information tasks, mainly in processes aimed at generating   (useful) entanglement 
between individually addressable parties, situated at spatially localized positions, 
has been primarily focused on systems of indistinguishable bosons \cite{killoran_2014, Morris_2020}, but has also been addressed in systems of indistinguishable fermions \cite{bouvrie_2017_b,bouvrie_2019}. 
Recently, an ensemble of ultracold identical particles has been splitted, and entanglement between the two spatially separated clouds was experimentally achieved in \cite{Lange_2018}, thus allowing for 
harnessing entangled states of many indistinguishable-particle systems for quantum information applications. 
%
Further, the distinction between correlations rooted at the antisymmetrization of the fermionic wave function and those arising from the interaction between particles has been explored  in systems of trapped atoms \cite{Becher_2020}. 

Both the study of  the generation of entanglement between (spatially localized) modes and between indistinguishable particles has been mainly focused on bipartite correlations.
In particular, a protocol based on a splitting-plus-detection operation has been proposed \cite{bouvrie_2017_b},  
that generates fermionic entanglement between a pair of identical fermions in a two-well potential out of an initial state with minimal (exchange) correlations.
As a result of the generation of entanglement between the particles ---created along the detection process---, final states can be prepared that are effective entangled states of two distinguishable qubits. 
In the present work we extend such protocol in pursuit of schemes that allow for the generation of \emph{multipartite} entangled states in systems of identical fermions, that ultimately become accessible multiqubit entangled states. 
Specifically, we consider a 3-fermion system that is amenable to be mapped into a 3-well potential, and 
explore the  possibility of generating fermionic 
\textsc{ghz}-type and \textsc{w}-type states 
 out of an initial stated endowed with exchange correlations only, by means of a tunneling-plus-particle detection process. 
 The fermionic 
\textsc{ghz}- and \textsc{w}-type states are the fermionic version of the standard \textsc{ghz}- and \textsc{w}-type states, defined in systems of three distinguishable qubits.
These pertain to different families of  3-qubit states, whose elements exhibit inequivalent kinds of  genuine entanglement (meaning that non-vanishing correlations exists across all the bipartitions of the system): the first (\textsc{ghz}) family comprises states which posses three-way entanglement \cite{Coffman2000},
whereas the second (\textsc{w}) family corresponds to states in which all the entanglement across any bipartition decomposes as sums of pairwise correlations, i.e., exhibit only two-way entanglement. 
We also explore whether the fermionic
 states resulting from the tunneling-plus-particle detection protocol can be effectively reduced to the usual 3-qubit \textsc{ghz}-type and \textsc{w}-type states, under certain cirmcumstances.


The work is structured as follows. 
In Section II we review some measures of multipartite entanglement defined for distinguishable-particle systems, as well as their `fermionic' counterpart, which are analogous entanglement measures applicable to indistinguishable-fermion systems. 
We also review the  protocol advanced in \cite{bouvrie_2017_b} for generating entanglement in a two-indistinguishable fermion system. 
After this preliminary section, we propose an extended protocol aimed at generating tripartite entanglement in a 3-fermion system (Section III), with emphasis on the generation of \textsc{ghz}-type states and highly entangled \textsc{w}-type states. 
In this latter case, the tunneling matrix that guarantees the preparation of a \textsc{w}-type state with maximal probability is explicitly given. 
A discussion related to the amount of useful (accessible) entanglement generated in the process is presented in Section IV. Finally, some conclusions are drawn in Section V.


\section{Preliminaries}

\subsection{Multipartite entanglement in distinguishable-parties systems}

A well-known
measure of \emph{bipartite} entanglement between two distinguishable parties ($A$ and $B$) in a pure state $\ket{\psi_{AB}}$ 
is given by the \emph{concurrence}
\begin{equation}
\label{bipConcurrence}
C_2(\ket{\psi_{AB}})=\sqrt{2(1-\textrm{Tr}\rho_A^2)}=\sqrt{2(1-\textrm{Tr}\rho_B^2)},
\end{equation}
where $\rho_{A(B)}=\textrm{Tr}_{B(A)}\ket{\psi_{AB}}\!\bra{\psi_{AB}}$ stands for the reduced density matrix of the subsystem $A (B)$. 
A suitable generalization of Eq. (\ref{bipConcurrence}) for a system of $N$ distinguishable parties in a pure state $\ket{\psi_N}$ is provided by the \emph{multipartite concurrence} \cite{Carvalho2004}
\begin{equation}
\label{NConcu}
C_N(\ket{\psi_N})=2^{1-(N/2)}\sqrt{(2^{N}-2)-\textrm{Tr}\sum_k\rho_k^2},
\end{equation}
 where $k$ labels \emph{all} the different reduced density matrices.
This quantity captures the entanglement present across all the bipartitions of the system. 
It vanishes for fully separable states $\ket{\psi_{N}}=\otimes^{N}_{i=1}\ket{\phi_{i}}$ and reaches its maximum value for \textsc{ghz} states $\ket{\psi_{N}}=\sum_{i=0}^1\ket{i...i}/\sqrt{2}$.
However, in multipartite systems, inequivalent types of multipartite entanglement may appear \cite{dur2000,verstraete2002}, and consequently different measures of entanglement may be considered. 
In particular, in a 3-qubit ($A$, $B$, and $C$) pure state, a 3-way entanglement arises quantified by the so-called 3-tangle $\tau$, defined as \cite{Coffman2000} 
\begin{equation}\label{3tangle} 
		\tau = \tau_{A(BC)} - \tau_{AB} - \tau_{AC},	
 \end{equation}
where $\tau_{ij}$ is the \emph{tangle} (squared concurrence), that accounts for the entanglement between subsystems $i$ and $j$. 
The 3-tangle distinguishes between two families of tripartite entangled 3-qubit pure states: the \textsc{ghz}-type family comprising the states with $\tau\neq 0$, and the \textsc{w}-type family, which includes states that are entangled across all bipartitions yet posses null 3-tangle \cite{dur2000}. 
In this way $\tau$, unlike $C_3$, detects only a specific type of tripartite entanglement (namely 3-way, \textsc{ghz}-type correlations), whereas $C_{3}$ captures all one-to-one bipartite correlations among the subsystems. 

\subsection{Multipartite entanglement in indistinguishable-fermion systems}

In composites of indistinguishable fermions, the antisymmetry of the state vectors prevents the fermionic states from being factorizable into products of single-particle states, giving rise to the exchange correlations, fundamentally encoded in Slater determinants (said to have Slater rank 1). 
In first quantization, the Slater determinant describing a $N$-fermion state in which the single-particles states $\{\ket{i_1},\dots,\ket{i_N}\}$ are occupied, reads
\begin{equation}
    \begin{split}
\ket{\psi^{sl}_{i_1...i_N}}&=\frac{1}{\sqrt{N!}}\sum_{j_1,...,j_N}\delta^{i_1...i_N}_{j_1...j_N} \ket{j_1}\otimes...\otimes\ket{j_N}\\
&=\hat {\mathcal A}\big( \ket{i_1}\otimes...\otimes\ket{i_N}\big),
    \end{split}
\end{equation}
where $\delta^{i_1..i_N}_{j_1...j_N}$ is the generalized Kronecker delta
\footnote{The Kronecker delta is defined as follows: $\delta^{i_1... i_N}_{j_1 ... j_N}=1$ if the indices $\{i_r\}$ are all different and ${i_1... i_N}$ is an even permutation of ${j_1... j_N}$;  $\delta^{i_1... i_N}_{j_1 ... j_N}=-1$ if the indices $\{i_r\}$ are all different and ${i_1... i_N}$ is an odd permutation of ${j_1... j_N}$; and $\delta^{i_1... i_N}_{j_1 ... j_N}=0$ in all other cases.}
, and $\hat {\mathcal A}$ the antisymmetrizer operator. 

As a result of the antisymmetry demand, the notion of entanglement between indistinguishable  particles (not between modes) has been reformulated \cite{schielmann_2001,Eckert_2002,Ghirardi_2002,plastino_2009_epl,tichy_2011_JPB,majtey_2016,bouvrie_2017_b,debarba2019}. 
Throughout this paper we will adhere to the definition of entanglement according to which 
pure states that are incompatible with the existence of a set of well-defined properties of all the individual subsystems will be regarded as entangled, or non-separable. Equivalently,  separable, or non-entangled states, are those for which a set of well-defined properties of the particles exists (see Ref. \cite{Majtey2023} for an introduction on this subject).
When this definition is applied to pure states of distinguishable parties, the standard definition of entanglement ---as correlations encoded in the non-factorizability of the state--- is recovered. 
Yet, when applied to pure states of indistinguishable  fermions, it turns out that, mathematically, 
the only states that are entangled are those that cannot be written as a single Slater determinant. 
In other words, 
the minimal (exchange) correlations do not contribute to the state’s entanglement, so a state $\ket{\psi_{N_f}}$ of $N$ indistinguishable fermions is regarded as non-entangled if and only if it can be expressed as a single Slater determinant. 
Otherwise, the state is said to exhibit \emph{fermionic entanglement}. 
 

In this context, the fermionic multipartite concurrence, analogous to (\ref{NConcu}) and valid for a pure $N$-fermion state, has been advanced in \cite{majtey_2016} as
\begin{equation}\label{concuf}
    	C_{N_f}(|\psi_{N_f}\rangle) = \sqrt{\alpha_N \left[N-1 - \sum_{n=1}^{N-1}\binom{N}{n} \textrm{Tr}\left(\rho_n^2\right)\right]},
\end{equation}
where $\rho_n$ stands for the reduced density matrix of the subsystem containing $n$ fermions, and $\alpha_N$ is the normalization constant
\begin{equation}
\alpha_N=\frac{1}{N-1-\sum_{n=1}^{N-1}\binom{N}{n}\binom{d}{\min[n,N-n]}^{-1}},
\end{equation}
being $d$ the dimension of the single-fermion Hilbert space $\mathcal H_f$. 

The 3-tangle $\tau$ has also been generalized  to 3-fermion pure states 
$\ket{\psi_{3_f}}$, provided $d=6$ \cite{Levay2008}. This generalization, to which we will refer to as 3-fermionic tangle, is given by
\begin{equation}\label{3tanglef}
 \begin{split}
         \tau_f (|\psi_{3_f}\rangle)&= 4 \,\Big|\!\left[\textrm{Tr}(AB)-P_{123}P_{456}\right]^2-4 \textrm{Tr}(A^{\#}B^{\#})\\
            &+ 4 P_{123} \det A + 4 P_{456}\det B\Big|,
        \end{split}
	\end{equation}
where $M^{\#}$ denotes the adjugate of the matrix $M$ \footnote{The adjugate of a matrix $M$ is the transpose of its cofactor matrix $C$, the latter defined as the matrix whose $(i, j)$ entry is the $(j, i)$ cofactor of $M$.}, and $A$ and $B$ are $3\times3$ matrices defined as
\begin{equation*}
A=\left(\begin{array}{ c c c }
			P_{156} & P_{164} & P_{145}\\
			P_{256} & P_{264} & P_{245}\\
			P_{356} & P_{364} & P_{345}
		\end{array}\right), \ \  B=\left(\begin{array}{ c c c }
			P_{423} & P_{431} & P_{412}\\
			P_{523} & P_{531} & P_{512}\\
			P_{623} & P_{631} & P_{612}
	\end{array}\right) ,
	\end{equation*}
with $P_{ijk}$ the coefficients in the expansion  
\begin{equation}\label{exp}
|\psi_{3_f}\rangle = \frac{1}{6}\sum_{i,j,k=1}^6 P_{ijk}
\ket{\psi_{ijk}^{sl}}.
\end{equation}

\subsection{Generation of fermionic entanglement in a two-fermion system}
In \cite{bouvrie_2017_b} a two-electron system in a double-well potential is considered. 
In addition to the two-level internal degree of freedom with states $\ket{\sdo}$ and $\ket{\su}$, each particle has a two-dimensional spatial degree of freedom with states $\ket{a}$ and $\ket{b}$, corresponding to (mutually orthogonal) wave functions spatially localized in a left ($a$) and right ($b$) well-potential. 
An orthonormal basis of the ensuing four-dimensional single-particle Hilbert space $\mathcal{H}_f$ is thus 
\ben\label{is2}
\ket{1}&=&\ket{a\su},\quad \ket{2}=\ket{a\sdo},\nonumber \\
\ket{3}&=&\ket{b\su},\quad \ket{4}=\ket{b\sdo}.
\een

Considering the initial state 
\be \label{initState}
\ket{\psi_{\textrm{init}}}=\hat{\mathcal{A}}(\ket{2}\otimes\ket{1})=\hat{\mathcal{A}}\big(\ket{a\sdo}\otimes\ket{a\su}\big),
\ee
a splitting, or tunneling, transformation $\hat T=\hat T_f\otimes \hat T_f$ is performed, with $\hat T_f$ a unitary operator acting on $\mathcal{H}_f$ such that
\begin{eqnarray}
\label{unitary}
\hat T_f|a\,\sigma\rangle&=&\sqrt{1-p}\,|a\,\sigma\rangle+\sqrt{p}\,|b\,\sigma\rangle,
\end{eqnarray}
where $\sigma=\,\su,\sdo$. 
Direct calculation verifies that the state $\hat T|\psi_{\text{init}}\rangle$ is a single Slater determinant, confirming that entanglement is not created by the splitting operation. This is consistent with the fact that the transformation $\hat{T}$ is a local unitary operation. At a second stage, the state $\hat T|\psi_{\text{init}}\rangle$ is projected onto the state with one particle in each well, obtaining (after normalization) the final state
\ben
\label{final1}
\ket{\psi_{\text{final}}}&=&\frac{1}{2}\Big(|a\sdo\rangle\otimes|b\su\rangle-|b\su\rangle\otimes|a\sdo\rangle\nonumber\\
&&+|b\sdo\rangle\otimes|a\su\rangle-|a\su\rangle\otimes|b\sdo\rangle\Big).
\een
Unlike the initial Slater determinant $|\psi_{\text{init}}\rangle$, the final state 
possess a non-vanishing amount of fermionic entanglement,
originated in the particle-detection process (or rather, by the concomitant projective measurement). Such non-zero entanglement becomes useful (accessible) entanglement between two effective and distinguishable qubits whenever agents (e.g. Alice and Bob)  have access to the particle in their 
corresponding well ($a$ and $b$, respectively). 
This amounts to `freeze' the spatial degrees of freedom (for example, by raising the tunneling  barrier), and involves ascribing the basis $\mathcal B_A=\{\ket{a\su},\ket{a\sdo}\}$ to Alice and $\mathcal B_B=\{\ket{b\su},\ket{b\sdo}\}$ to Bob. 
Consequently, when Alice and Bob access the particle in their respective location, each of them posses a single qubit (distinguishable from the other one), and in line with the state (\ref{final1}) they share 
the maximally entangled two-qubit state:
\be
\label{projectedDisting}
\ket{\psi_q}=\frac{1}{\sqrt{2}}\Big(\ket{\sdo}_A\otimes \ket{\su}_B-\ket{\su}_A\otimes \ket{\sdo}_B\Big),
\ee
and \cite{bouvrie_2017_b}
\begin{equation}
C_2(\ket{\psi_q})=C_{2f}(\ket{\psi_{\textrm{final}}}).
\end{equation}
In this way, a maximally entangled state of two distinguishable (addressable) qubits can be generated by means of a tunneling plus a particle detection process, applied on a two indistinguishable fermion system.

\section{Generation of tripartite fermionic entanglement}
We now address the problem of generalizing the protocol just described ---that generates fermionic \emph{bipartite} entanglement out of Slater determinants---, to a protocol that creates fermionic \emph{multipartite} entanglement from an initially separable state. 
In particular, we focus on the generation of tripartite entanglement among three indistinguishable fermions, considering \emph{fermionic} \textsc{w}-type states (or $\textsc{w}_f$ for short) and \emph{fermionic} \textsc{ghz}-type states ($\textsc{ghz}_f$). 

The $\textsc{w}_f$ state is the fermionic analogue of the $\textsc{w}$-type state of 3 distinguishable qubits $\ket{\textsc{w}_q}=c_1\ket{001}+c_2\ket{010}+c_3\ket{001}$, and reads 
\ben\label{fw}
\ket{\textsc{w}_f}=c_1\ket{\psi^{sl}_{ijk}}+c_2\ket{\psi^{sl}_{ilm}}+c_3\ket{\psi^{sl}_{njl}},
 \een
where $i,j,k,l,m,n$ are different indices taken from the set of single-particle states $\{\ket{i}\}$, and $\sum^3_{s=1}|c_s|^2=1$. 
For non-vanishing, yet arbitrary, values of the coefficients $c_{s}$, $\ket{\textsc{w}_q}$ pertains to the \textsc{w} family of 3-qubit states, whose \emph{representative} element is that corresponding to $c_1=c_2=c_3=1/\sqrt{3}$ \cite{dur2000} (the resulting state is customarily called simply the W state).
$\ket{\textsc{w}_f}$ shares with $\ket{\textsc{w}_q}$ the property that under 
tracing over any of the parties the result is an entangled two-party (mixed) state \footnote{In line with the definition of fermionic entanglement given above, mixed separable states are those than can be expressed as a convex sum of pure states of Slater rank 1 \cite{Ghirardi_2002}. If such decomposition is not possible, the fermionic state is said to be entangled.}. 
Further, $\ket{\textsc{w}_f}$ has vanishing 3-fermionic tangle $\tau_f$, as $\ket{\textsc{w}_q}$ has vanishing $\tau$.

Similarly, the state $\textsc{ghz}_f$ is the fermionic analogue of the standard $\textsc{ghz}$-type state of 3 distinguishable qubits, given by  $\ket{\textsc{ghz}_q}=c_1\ket{000}+c_2\ket{111}$. 
The fermionic \textsc{ghz}-type state reads explicitly \cite{majtey_2016}
\ben\label{fghz}
\ket{\textsc{ghz}_f}
=c_1\ket{\psi^{sl}_{ijk}}+c_2\ket{\psi^{sl}_{lmn}},
\een  
where again all $i,j,k,l,m,n$ are different indices taken from the set of single-particle states $\{\ket{i}\}$, and 
$\sum^2_{s=1}|c_s|^2=1$. For arbitrary non-vanishing coefficients $c_s$, the state $\ket{\textsc{ghz}_q}$ pertains to the \textsc{ghz} family of 3-qubit states, whose paradigmatic element is that corresponding to $c_1=c_2=1/\sqrt{2}$ (known simply as the GHZ state) \cite{dur2000}.
For these values of the coefficients $c_s$, the state (\ref{fghz}) possess the maximum amount of multipartite fermionic concurrence $C_{3_f}$,
while the reduced two-fermion density matrices correspond to
separable mixed states. 
This property of a state
being maximally entangled, while tracing over any one of the
subsystems destroys any entanglement present, is characteristic of the GHZ state, and reinforces the analogy between $\ket{\textsc{ghz}_f}$ and $\ket{\textsc{ghz}_q}$. 
Finally, notice that the existence of the  $\textsc{w}_f$ and the  $\textsc{ghz}_f$ states requires a single-particle Hilbert space with $d\geq 6$.

\subsection{The protocol}

Consider a system of 3 indistinguishable fermions with a 6-dimensional single-particle Hilbert space $\mathcal{H}_f$. 
Each fermion possess a two-level internal degree of freedom with states $\{\ket{\sigma}\}=\{\ket{\su},\ket{\sdo}\}$, and a three-level spacial degree of freedom, with orthogonal states $\{\ket{S}\}=\{\ket{a},\ket{b},\ket{c}\}$, corresponding to spatially localized states in a 3-well potential with periodic boundary conditions (see for example \cite{henderson2009experimental} for the controlled generation of potentials with arbitrary geometries using a moving laser beam). 
An orthonormal basis of $\mathcal{H}_f$ is thus $\{\ket{i}=\ket{S\sigma}=\ket{S}\otimes\ket{\sigma}\}$, where the states are ordered according to 
\ben\label{is}
\ket{1}&=&\ket{a\su},\quad \ket{2}=\ket{a\sdo},\nonumber \\
\ket{3}&=&\ket{b\su},\quad \ket{4}=\ket{b\sdo},\\
\ket{5}&=&\ket{c\su},\quad \ket{6}=\ket{c\sdo}.\nonumber
\een

An initial Slater determinant $\ket{\psi_{\textrm{init}}}=\mathcal{\hat A} (\ket{i}\otimes\ket{j}\otimes\ket{k})$ with only two wells populated
is let to evolve under a unitary evolution 
\be \label{unitaria}
\hat U=\hat U_f\otimes \hat U_f\otimes \hat U_f,
\ee
with $\hat U_f$ acting on $\mathcal H_f$, and afterwards is projected onto the subspace spanned by the states having one particle in each well. 
This last step is necessary in order to  
create states that transform into accessible tripartite entangled states when `freezing' the spatial degree of freedom, i.e., when effectively distinguishing the particles that become accessible to three separate agents 
Alice, Bob, and Charlie (in sites $a,b$ and $c$, respectively), in a way analogous to that explained just before Eq. (\ref{projectedDisting}) (see also Section \ref{freezing}).

In order to have a non-zero projection onto the subspace of states with one fermion in each well, we allow the particles to tunnel between neighboring sites during a finite time, so the effective unitary splitting transformation (acting on the spatial degrees of freedom) reads
\be\label{split}
\hat T=\hat T_f\otimes \hat T_f\otimes \hat T_f,
\ee
where the matrix representation of $\hat T_f$ in the basis $\{\ket{S}\}$ is given by 
\be
\label{tunnel}
T =\left(\begin{array}{ c c c }
            t_{aa} & t_{ab} & t_{ac}\\
            t_{ba} & t_{bb} & t_{bc}\\
            t_{ca} & t_{cb} & t_{cc}
        \end{array}\right),
\ee
with $t_{\alpha\beta}\equiv\langle \alpha|T|\beta\rangle$ denoting the probability amplitude of the transition $\beta\rightarrow \alpha$ ($\alpha,\beta\in \{a,b,c\}$), so accordingly $|t_{\alpha\beta}|^2$ stands for the probability of the corresponding transition 
(below we will focus on
the symmetric case $|t_{\alpha\beta}|^2=|t_{\beta\alpha}|^2$, which implies that the  probability of tunneling between sites $\alpha$ and $\beta$ does not depend on the direction of the transition).
We will also allow for superpositions of vectors with common spatial states but opposite (flipped) internal states. 
Since $\ket{\sigma}$ is invariant under $\hat T$, an additional spin-flip operation is therefore introduced via a unitary transformation (acting on the spinorial degrees of freedom)
\be\label{flip}
\hat \Sigma=\hat \Sigma_f\otimes \hat \Sigma_f\otimes \hat \Sigma_f,
\ee
whose matrix representation, in the basis $\{\ket{\sigma}\}$, is 
\be
\Sigma =\left(\begin{array}{ c c }
            s_{\uparrow \uparrow } & s_{\uparrow \downarrow }\\
            s_{\downarrow \uparrow } & s_{\downarrow \downarrow }
        \end{array}\right), 
\ee
where $s_{\sigma\sigma^\prime}\equiv\langle \sigma|\Sigma|\sigma'\rangle$ denotes the probability amplitude of the transition $\sigma'\rightarrow \sigma$ (and $|s_{\sigma\sigma^\prime}|^2$ therefore stands for the corresponding transition probability).
In this way, the single-particle  unitary operator $\hat U_f$ in (\ref{unitaria})
decomposes as
\begin{equation}\label{singleU}
\hat U_f=\hat T_f\otimes \hat \Sigma_f.
\end{equation}

Since  
$\hat U$ acts locally on each single-particle Hilbert space, it does not create fermionic entanglement. Consequently, the (unitarily) evolved state is a Slater determinant, i.e.,
\ben\label{evolved}
\ket{\psi}=\hat U\ket{\psi_{\textrm{init}}}&=&\hat U \hat {\mathcal A}(\ket{i}\otimes\ket{j}\otimes\ket{k})\\\nonumber
&=&\hat {\mathcal A}(\ket{i'}\otimes\ket{j'}\otimes\ket{k'}),
\een
with $\ket{i'}=\hat U_f\ket{i}$.
As discussed in \cite{bouvrie_2017_b}, the generation of fermionic entanglement arises in the particle-detection process, when $\hat U\ket{\psi_{\textrm{init}}}$ is projected onto the subspace of states with no empty wells.

\subsection{Generation of GHZ- and W-type fermionic states}

 Without loss of generality we may assume that the initial state with only two wells populated is 
 \be \label{inicialb}
\ket{\psi_{\textrm{init}}}=\hat{\mathcal{A}}(\ket{1}\otimes\ket{3}\otimes\ket{4})=\hat{\mathcal{A}}\big(\ket{a\su}\otimes\ket{b\su}\otimes\ket{b\sdo}\big),
\ee
so the evolved state (\ref{evolved}) reads 
\begin{eqnarray}\label{psievol}
\ket{\psi}&=&\sum _{\alpha \beta \gamma \ } \sum _{\sigma \sigma '\sigma ''\ } t_{\alpha a} t_{\beta b} t_{\gamma b} \ s_{\sigma \uparrow } s_{\sigma '\uparrow } s_{\sigma ''\downarrow } \times \nonumber\\ &&\hat{\mathcal{A}} \big( \ket{\alpha \,\sigma } \otimes \ket{\beta \,\sigma '} \otimes \ket{\gamma \,\sigma ''}\big) ,
    \end{eqnarray}
where $\alpha,\beta,\gamma\in\{a,b,c\}$, and $\sigma,\sigma',\sigma''\in\{\uparrow,\downarrow\}$.
\begin{figure}[ht]
\includegraphics[width=0.8\columnwidth]{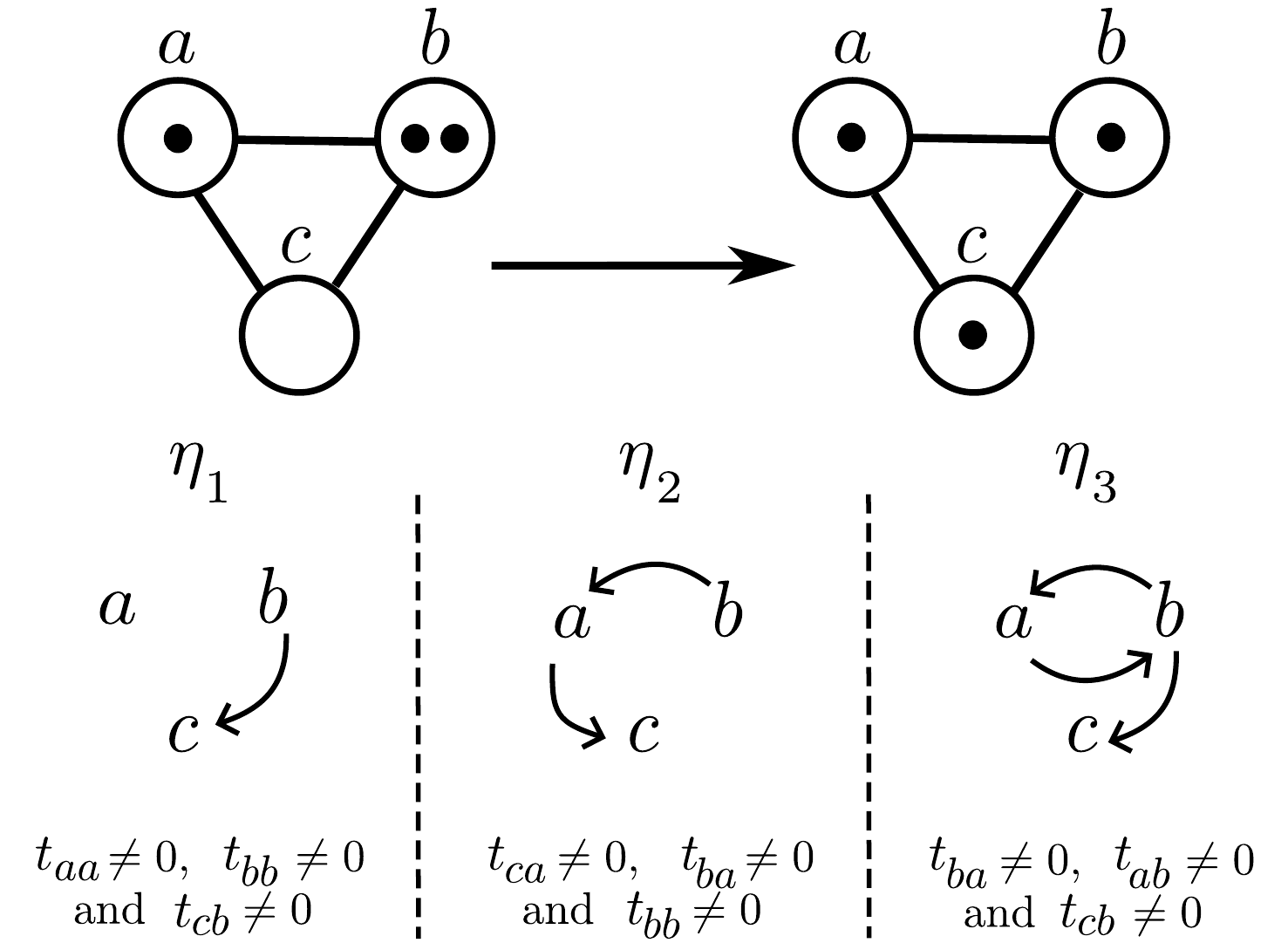}
\caption{Upper panel: schematic representation of the spatially localized sites $a,b$ and $c$, with periodic boundary conditions. Initially the wells $a$ and $b$ are populated, and then the particles can tunnel between neighboring sites, so all three wells will be occupied with certain probability. Bottom panel: The parameter $\eta_{i}$ indicates that $i$ particles ($i=1,2,3$) have tunneled between adjacent sites.}
\label{trans}
\end{figure}
For simplicity in the writing we introduce the following notation 
\be
 \ket{\sigma_a \sigma '_b\sigma ''_c}^{-} \equiv\ \hat{\mathcal{A}} \big( \ket{a\,\sigma } \otimes \ket{b\,\sigma '} \otimes \ket{c\,\sigma ''}\big) ,
\ee
and define the parameters 
\begin{eqnarray}\label{defzetas}
z_{1} &=&s_{\uparrow \uparrow }^{2} s_{\uparrow \downarrow } ,\quad z_{2} =s_{\downarrow \uparrow }^{2} s_{\downarrow \downarrow } ,\nonumber\\
z_{3} &=&s_{\uparrow \uparrow } s_{\uparrow \downarrow } s_{\downarrow \uparrow } ,\quad z_{4} =s_{\downarrow \downarrow } s_{\downarrow \uparrow } s_{\uparrow \uparrow } ,\\
z_{5} &=&s_{\downarrow \uparrow }^{2} s_{\uparrow \downarrow } ,\quad z_{6} =s_{\uparrow \uparrow }^{2} s_{\downarrow \downarrow } \nonumber,
\end{eqnarray}
and 
 \begin{eqnarray}\label{etas}
        \eta_1&=&t_{aa} t_{bb} t_{cb} ,\nonumber\\ 
        \eta_2&=&t_{ca} t_{ab} t_{bb} ,\\
        \eta_3&=&t_{ba} t_{ab} t_{cb}.\nonumber 
    \end{eqnarray}    
Notice that $\eta _{1}$ stands for the probability amplitude of  a \emph{single} particle tunneling between sites (in this case a fermion hops from $b$ to $c$ whereas the remaining two stay in their initial well). 
Analogously, $\eta _{2}$ is the probability amplitude for \emph{two} fermions tunneling between sites ($b\rightarrow a$, and $a\rightarrow c$), whereas $\eta _{3}$ stands for the probability amplitude for \emph{three} fermions tunneling between sites ($a\rightarrow b$, $b\rightarrow a$ and $b\rightarrow c$). 
Figure \ref{trans} depicts the disposition of the wells, and the interpretation just described of the parameters $\eta_i$. 

With this notation, after some lenghtly but direct calculation we can rewrite (\ref{psievol}) as
\begin{widetext}
\begin{eqnarray}\label{psicomp}
\ket{\psi} &=&\left(\eta_{3}-\eta_{2}\right)\left[\left(z_6-z_3\right)\ket{\downarrow_a\uparrow_b\uparrow_c}^{-}+\left(z_5-z_4\right)\ket{\uparrow_a\downarrow_b\downarrow_c}^{-}\right]+\left(\eta_{2}-\eta_{1}\right)\left[\left(z_6-z_3\right)\ket{\uparrow_a\downarrow_b\uparrow_c}^{-}+\left(z_5-z_4\right)\ket{\downarrow_a\uparrow_b\downarrow_c}^{-}\right]\nonumber\\
    &&+\left(\eta_{1}-\eta_{3}\right)\left[\left(z_6-z_3\right)\ket{\uparrow_a\uparrow_b\downarrow_c}^{-}+\left(z_5-z_4\right)\ket{\downarrow_a\downarrow_b\uparrow_c}^{-}\right] +\,\textrm{``terms with one empty well"}\nonumber\\
&=&\left(\eta_{3}-\eta_{2}\right)\left[\left(z_6-z_3\right)\ket{\psi^{sl}_{235}}+\left(z_5-z_4\right)\ket{\psi^{sl}_{146}}\right]+\left(\eta_{2}-\eta_{1}\right)\left[\left(z_6-z_3\right)\ket{\psi^{sl}_{145}}+\left(z_5-z_4\right)\ket{\psi^{sl}_{236}}\right]\nonumber\\
    &&+\left(\eta_{1}-\eta_{3}\right)\left[\left(z_6-z_3\right)\ket{\psi^{sl}_{136}}+\left(z_5-z_4\right)\ket{\psi^{sl}_{245}}\right] +\,\textrm{``terms with one empty well"}.
    \end{eqnarray}
  
\end{widetext}

  Comparison of this last expression with (\ref{fghz}) shows that in order to obtain a $\textsc{ghz}_f$ state \emph{after detection of one particle in each well} (that is, after discarding the contributions with one empty well), it is necessary that one and only one of the terms enclosed within square brackets  is non-zero. 
However, having two vanishing coefficients, so $\eta_{n}-\eta_{m}=0=\eta_{m}-\eta_{k}$ (with $n,m,k \in\{1,2,3\}$) implies that all $\eta_n$ are equal, hence we cannot have only one non-zero term among $\eta_{3}-\eta_{2}$, $\eta_{2}-\eta_{1}$ and $\eta_{1}-\eta_{3}$. 
Consequently, \emph{it is not possible} to obtain a $\textsc{ghz}_f$ state by means of the proposed process.

We now focus on the conditions required for generating a $\textsc{w}_f$ state. 
Inspection of (\ref{psicomp}) and (\ref{fw}) shows that a state $\ket{\textsc{w}_f}$ may be created once a particle has been detected in each well, provided either one of the following 
conditions is met:
\begin{subequations}
\begin{eqnarray}
z_6-z_3&=&s_{\su\su}\det \Sigma=0,\\
z_4-z_5&=&s_{\sdo \su}\det \Sigma=0.
\end{eqnarray}
\end{subequations}
Since $\Sigma$ is a unitary matrix (so $|\!\det \Sigma|=1$), these conditions amount to take either $s_{\su\su}=0$ or $s_{\sdo\su}=0$.
By taking the latter case
it follows that the probability $|s_{\sdo\su}|^2$ is null, and by virtue of the normalization of the transition probabilities, if the probability of flipping the states $\ket{\su},\ket{\sdo}$ is zero, the probability of remaining in the same state equals 1, meaning that no flipping operation is indeed required (so $\Sigma=\mathbb I$), and the tunneling plus particle detection process suffice in principle to generate a fermionic \textsc{w}-type state.

With the above considerations, (\ref{psicomp}) reduces to 
\begin{equation}\label{prepretipow}
\ket{\psi} =\sqrt{\mathcal P}\ket{\psi_{\textsc{w}}}
+\textrm{``terms with one empty well"},
\end{equation}
where
\begin{equation}\label{P}
\mathcal P =|\eta_3-\eta_2|^2+|\eta_2-\eta_1|^2+|\eta_1-\eta_3|^2,
\end{equation}
and $\ket{\psi_{\textsc{w}}}$ is the normalized state
\begin{eqnarray}\label{tipow}
\ket{\psi_{\textsc{w}}} &=&\frac{1}{\sqrt{\mathcal P}}\Big[( \eta _{3} -\eta _{2})\ket{\psi^{sl}_{235} } +( \eta _{2} -\eta _{1})\ket{\psi^{sl}_{145} } \nonumber\\
    &&+( \eta _{1} -\eta _{3})\ket{\psi^{sl}_{136} }\Big].
\end{eqnarray}
Once the system is in the state (\ref{prepretipow}), a detection of particles in all three wells will  occur with probability $\mathcal P$. 
If the detection takes place, the ensuing state ---obtained by projecting (\ref{prepretipow}) onto the subspace spanned by states with all three sites occupied--- is 
\begin{equation}\label{final}
\ket{\psi _{\textrm{final}}} =\ket{\psi_{\textsc{w}}},
\end{equation}
which has the structure of a $\textsc{w}_f$ state. Its fermionic concurrence can be directly computed from (\ref{concuf}), and reads
\begin{equation}\label{concus}
 C_{3f} (\ket{\psi_{\textrm{final}}})= \sqrt{\frac{4}{3}\Big(
 1-\sum^3_{i=1}|r_i|^4
\Big)},
\end{equation}
where $|r_i|=|\eta_j-\eta_k|/\sqrt{\mathcal P}$ (with $i\neq j \neq k$).


\subsection{Generation of highly entangled fermionic $W$-type states}

By maximizing the term within parenthesis in Eq. (\ref{concus}) (subject to the constriction  $\sum_i|r_i|^2=1$), it follows that the entanglement of the final state is maximal whenever all the $|r_i|$'s are equal, that is, whenever 
 %
\begin{equation}\label{eq:equalcoefW}
 |\eta_3-\eta_2|= |\eta_2-\eta_1|= |\eta_1-\eta_3|.
 \end{equation}
To disclose the structure of the matrices $T$ that guarantee this condition, we 
resort to the parametrization of $SU(3)$ based on Euler angles given in \cite{Byrd1998,Byrd2000,Cveti2002,Tilma_2002}. 
Following it, the tunneling matrix can be written as 
\begin{equation}\label{Tmat}
    T=e^{i \theta_1 \lambda_3}e^{i \theta_2 \lambda_2}e^{i \theta_3 \lambda_3}e^{i \theta_4 \lambda_5}e^{i \theta_5 \lambda_3}e^{i \theta_6 \lambda_2}e^{i \theta_7 \lambda_3}e^{i \theta_8 \lambda_8},
    \end{equation}
where $\lambda_l$ (with $l=1, ..., 8)$ stand for the Gell-Mann matrices, and $\theta_l$ are real parameters with $\theta_1$, $\theta_3$, $\theta_5, \theta_7 \in [0,\pi)$, $\theta_2, \theta_4, \theta_6 \in[0,\pi/2]$, and $\theta_8 \in[0,2\pi)$ \cite{byrd1997geometry}.
The condition (\ref{eq:equalcoefW}) restricts the allowed values of the  $\theta_l$, according to the couple of equations (see the Appendix \ref{equalcoef})  
\begin{subequations}\label{cond}
\begin{eqnarray}
\cos \theta_4 &=& -\cos \theta\, \tan(2 \theta_2)\,\cot\theta_6,\label{condb}\\
\cos ^2\theta &=&  \frac{\cos ^2(2 \theta_2 )}{3+\cos ^2(2 \theta_2 )},\label{conda}
\end{eqnarray}
\end{subequations}
with 
\begin{equation}\label{sum}
\theta=2\,(\theta_{3}+\theta_{5}).
\end{equation}

The solutions of (\ref{cond})
can be simplified by focusing on tunneling  matrices for which $|t_{\alpha\beta}|^2=|t_{\beta\alpha}|^2$, meaning that the probability of tunneling from site $\alpha$ to $\beta$ equals the probability of hopping from site $\beta$ to $\alpha$.
As shown in the Appendix \ref{equalcoef}, this symmetry condition implies that 
\ben
\theta_2 = \theta_6, \quad \theta_1 = \theta_5=\theta_7 = \theta_8=0.
\een

Gathering results, the tunneling matrices $\tilde{T}=[\tilde{t}_{\alpha\beta}]$ such that $|\tilde t_{\alpha\beta}|^2=|\tilde t_{\beta\alpha}|^2$, that produce 
$\textsc{w}_f$ states consistent with (\ref{eq:equalcoefW}), factorize as   
\begin{equation}
\tilde{T}=e^{i \theta_2 \lambda_2}
e^{i \theta_3 \lambda_3}
e^{i \theta_4 \lambda_5}e^{i \theta_2 \lambda_2},
 \end{equation}
where the parameters $\theta_2, \theta_4$, and $\theta=2\theta_3$ satisfy 
\begin{subequations}\label{condbis}
\begin{eqnarray}
\cos \theta_4 &=& -\frac{2\cos 2\theta_3}{1-\tan^2 \theta_2},\\
\cos^2 (2\theta_3) &=& \frac{\cos^2 (2 \theta_2)}{3 + \cos^2 (2 \theta_2)}.
\end{eqnarray}
\end{subequations}

Resorting to the explicit form of the Gell-Mann matrices, $\tilde{T}$ rewrites as
\begin{equation}\label{Tsim}
\tilde{T}=R_z(\theta_2) \,e^{i \theta_3 \lambda_3}\,R_y(\theta_4)\, R_z(\theta_2),
\end{equation}
where $R_z(\theta_2)=e^{i\theta_2 \lambda_2}$ corresponds to a rotation by an angle $-\theta_2$ around the $z$-axis,
and $R_y(\theta_4)=e^{i\theta_4 \lambda_5}$ to 
a rotation by an angle $\theta_4$ around the $y$-axis. 
Only for $\theta_3=0$ the tunneling operator $\tilde{T}$ reduces to a product of rotations. 
In such case, however, the first condition in (\ref{condbis}) cannot be met, and consequently no solution exists for which $\tilde{T}$ reduces to a product of rotations $R_z(\theta_2) \,R_y(\theta_4)\, R_z(\theta_2)$.
 \begin{figure}[]
\centering
\includegraphics[width=0.85\linewidth]{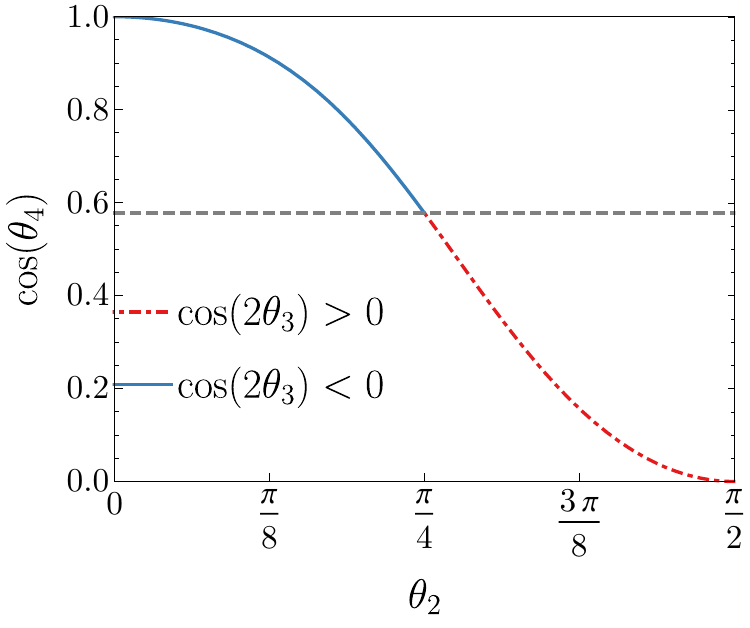}
\caption{Allowed values for $\cos \theta_4$ in the symmetric case $|t_{\alpha\beta}|=|t_{\beta\alpha}|$, as a function of the free parameter $\theta_2$. The grey dashed line corresponds to $\cos \theta_{4}= 1/\sqrt{3}$.}
\label{fig:gell-mannsimetrico}
\end{figure}

The phase $\theta_3$ can be eliminated from the pair of equations (\ref{condbis}), resulting in the single condition 
\begin{equation}\label{cos4}
|\cos\theta_4|=\cos\theta_4=\frac{2\,\cos^2\theta_2}{\sqrt{3+\cos^2(2\theta_2)}},
\end{equation}
where the first equality follows from fact that $0<\theta_4<\pi/2$ (the values $\theta_2,\theta_4=0,\pi/2$ are excluded, as discussed in the Appendix \ref{equalcoef}).
This leaves us with a single free parameter, $\theta_2\in(0,\pi/2)$.
Figure \ref{fig:gell-mannsimetrico} shows the allowed values of $\theta_4$ for varying values of $\theta_2$.

\begin{figure}[]
\centering
\includegraphics[width=0.85\linewidth]{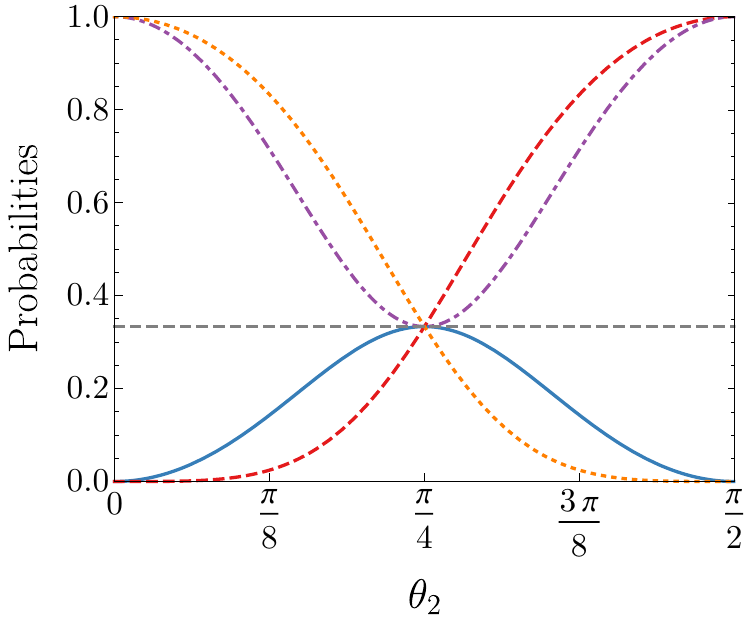}
\caption{Tunneling probabilities: $|\tilde t_{ab}|^2= |\tilde t_{ac}|^2$ (solid blue line), and $|\tilde t_{bc}|^2$ (dashed red line). 
Permanence probabilities: $|\tilde t_{aa}|^2$ (dotted-dashed purple line), and $|\tilde t_{bb}|^2=|\tilde t_{cc}|^2$ (dotted orange line).
All the probabilities coincide, and are equal to $1/3$, at $\theta_2 = \pi/4$. }
\label{fig:gell-manntunnelings}
\end{figure}

From Eqs. (\ref{Tsim}) and  (\ref{tunnel}) we can compute the elements $\tilde t_{\alpha\beta}$ of $\tilde T$, whose explicit expressions are shown in Eq. (\ref{eq:tunnelings}). From that, and by
using conditions (\ref{condbis}), direct calculation gives the following transition probabilities 
\begin{subequations}
\begin{eqnarray}
|\tilde t_{ac}|^2&=&\cos^2\theta_2 \sin^2\theta_4\nonumber\\
&=&\frac{\sin^2(2\theta_2)}{4-\sin^2(2\theta_2)}=
|\tilde t_{ab}|^2,\\
|\tilde t_{bc}|^2&=&\sin^2\theta_2 \sin^2\theta_4=\frac{\sin^4\theta_2}{1-\sin^2\theta_2\cos^2\theta_2},
\end{eqnarray}
\end{subequations}
which are respectively shown in the blue (solid) and red (dashed) lines in Figure \ref{fig:gell-manntunnelings}, as a function of $\theta_2$. 
As for the permanence probabilities, we get 
\begin{subequations}
\begin{eqnarray}
|\tilde t_{bb}|^2&=&|\tilde t_{cc}|^2=\cos^2\theta_4=\frac{8 \cos ^4\theta_2}{7+\cos (4 \theta_2)},\\
|\tilde t_{aa}|^2&=&\frac{5+3 \cos (4 \theta_2)}{7+\cos (4 \theta_2)},
\end{eqnarray}
\end{subequations}
respectively plotted in the orange (dotted) and the purple (dotted-dashed) curves in Fig. \ref{fig:gell-manntunnelings}.
Clearly, for $\theta_2=\pi/4$  (meaning $\cos (2\theta_3)=0$ and $\cos \theta_4=1/\sqrt{3}$, as follows from Fig. \ref{fig:gell-mannsimetrico}), all the tunneling and permanence probabilities coincide, and are equal to $1/3$. 
The corresponding  tunneling matrix and final state thus read (with $\theta_3=\pi/4$) 
\begin{equation}\label{eq:Tmatrixopt}
\tilde T=\frac{1}{\sqrt{3}}\left(
    \begin{array}{ccc}
    e^{i\frac{ 7 \pi }{12}} & e^{-i\frac{\pi}{12} } & e^{i\frac{ \pi }{4}} \\
    -e^{-i\frac{ \pi }{12}} & e^{-i\frac{5 \pi}{12}   } & e^{-i\frac{3\pi}{4}  } \\
     -1 & -1 & 1 \\
    \end{array}
    \right),
\end{equation}
and 
\ben \label{finalpart}
\ket{\psi_{\textrm{final}}}&=\frac{1}{\sqrt{3}}\left(e^{-i\frac{\pi}{3}}\ket{\psi^{sl}_{235}} +e^{i\frac{\pi}{3}}\ket{\psi^{sl}_{145}} -\ket{\psi^{sl}_{136}}\right).
\een
\begin{figure}[h]
 \centering
\includegraphics[width=0.85\columnwidth]{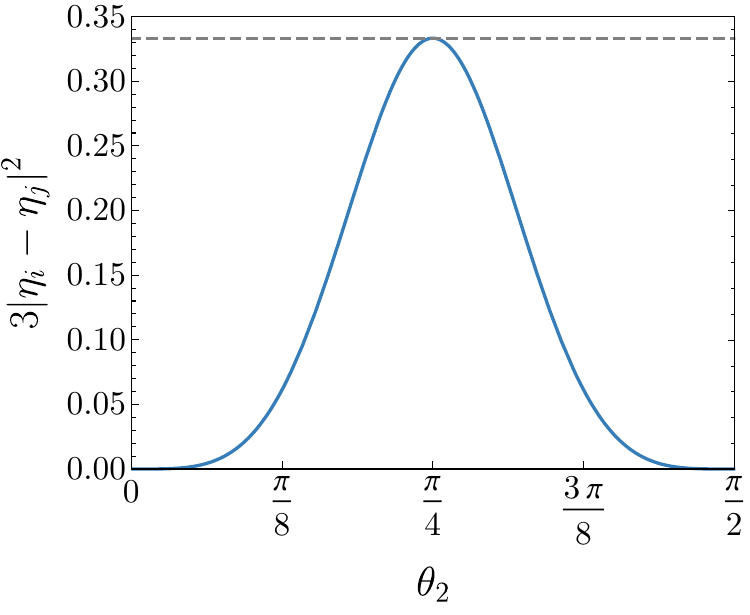}
\caption{Probability $\mathcal P$ of detection of one particle in each well for an equally-weighted superposition of the form (\ref{tipow}), assuming $|\tilde t_{\alpha\beta}|^2=|\tilde t_{\beta\alpha}|^2$. For $\theta_2=\pi/4$ the probability attains it maximum ($1/3$) value.}
 \label{prob}
\end{figure}

Further, as can be seen in Fig. \ref{prob}, the values of $\theta_2$ for which the transition probabilities coincide also maximize the probability (\ref{P}), which in the present case reduces to  
\begin{eqnarray}
\mathcal P&=&3|\eta_3-\eta_2|^2=3|\eta_2-\eta_1|^2=3|\eta_1-\eta_3|^2\nonumber\\
&=&3 \sin^2\theta_4 \cos^2\theta_4 \sin^2\theta_{2}\nonumber\\
&=&\frac{12 \sin ^4(2 \theta_2)}{[7+\cos \,(4 \theta_2)]^2},
\end{eqnarray}
where in the second line we resorted to Eq. (\ref{ape}) with $\theta_6=\theta_2$.
This means that the probability of having a fermionic w-type state increases as the tunneling probabilities tend to coincide. 

The simplest system in which all these conditions can be met, so the state (\ref{finalpart}) can be prepared with maximal probability, corresponds to an array of three identical potential wells, in which the onsite energies coincide (meaning that the pairs of states $\ket{1}$ and $\ket{2}$, $\ket{3}$ and $\ket{4}$, and $\ket{5}$ and $\ket{6}$ are degenerated). 
This guarantees that the condition $|\tilde t_{ab}|^2=|\tilde t_{bc}|^2=|\tilde t_{ca}|^2$ is satisfied. 

\section{Amount of accessible generated entanglement}\label{freezing}

Once the system is prepared in the final $\textsc{w}_f$-type state (\ref{final}), the spatial degrees of freedom may be `frozen', for example by rising the tunneling barrier, in such a way that the particles become accessible to localized agents in each well (Alice, Bob and Charlie in sites $a,b$ and $c$, respectively).
If we then ascribe the basis $\mathcal B_A=\{\ket{a\su},\ket{a\sdo}\}$ to Alice,  $\mathcal B_B=\{\ket{b\su},\ket{b\sdo}\}$ to Bob, and  $\mathcal B_C=\{\ket{c\su},\ket{c\sdo}\}$ to Charlie, so that each agent has access to a single two-level system (distinguishable from the remaining two), the originally indistinguishable-fermion state  (\ref{tipow}) becomes the effective 3-distinguishable qubit state 
\begin{align}\label{3qu}
\ket{\textsc{w}_q}=\frac{1}{\sqrt{\mathcal P}}&\Big[( \eta_{3} -\eta _{2})\ket{\downarrow}_A\otimes\ket{\uparrow}_B \otimes \ket{\uparrow}_C \nonumber\\
&+( \eta_{2} -\eta_{1})\ket{\uparrow}_A\otimes\ket{\downarrow}_B \otimes\ket{\uparrow }_C \nonumber\\
    &+( \eta_{1} -\eta _{3})\ket{\uparrow}_A\otimes \ket{\uparrow }_B\otimes\ket{\downarrow }_C\Big],
\end{align}
shared by the three parties.
Notably, the amount of fermionic entanglement $C_{3_f}$ of the final state (\ref{tipow}), given by (\ref{concus}), coincides with the amount of tripartite entanglement $C_3$ in the 3-qubit state (\ref{3qu}) \emph{after due normalization} 
(recall that the fermionic multipartite concurrence is already a normalized quantity, yet $C_N$ in Eq. (\ref{NConcu}) is not). That is, 
\begin{equation}\label{concufyq}
 C_{3f} (\ket{\psi_{\textsc{w}}})=\frac{C_3(\ket{\textsc{w}_q})}{C_3^\textrm{max}},
\end{equation}
where $C_3^\textrm{max}$ stands for the maximum value of $C_{N=3}$. 
Specifically,  $C_3^\textrm{max}=\sqrt{3/2}$, and is the maximum tripartite entanglement, attained for the GHZ state $(1/\sqrt{2})(|000\rangle+|111\rangle)$.  
For the particular (equally-weighted, symmetric) case in which the final state is given by (\ref{finalpart}), $|r_i|=|\eta_j-\eta_k|/\sqrt{\mathcal P}$ reduces to $|r_i|=1/\sqrt{3}$, and Eq. (\ref{concus}) gives 
\begin{equation}\label{concuspart}
 C_{3f} (\ket{\psi_{\textrm{final}}})= \sqrt{\frac{4}{3}\Big(
 1-\sum^3_{i=1}\frac{1}{9}
\Big)}=\frac{2\sqrt{2}}{3}\approx 0.94.
\end{equation}
For the tripartite concurrence $C_3$
it holds that \cite{Carvalho2004}
    \begin{equation}
    \frac{C_3(\ket{\textsc{w}_q})}{C_3(\ket{\textsc{ghz}_q})}=\frac{C_3(\ket{\textsc{w}_q})}{C_{3}^{\max}}=\frac{2\sqrt{2}}{3}
    \end{equation}
thus verifying Eq. (\ref{concufyq}).
 

As for the 3-tangle and its fermionic counterpart, we have that for triads of distinguishable qubits  $\tau(\ket{\textsc{w}_q}) = 0$. 
For a 3-fermion system in the state (\ref{tipow}), the only non-zero coefficients in the expansion (\ref{exp}) are $P_{235}, P_{145}, P_{136}$, whence direct calculation of the fermionic 3-tangle resorting to Eq. (\ref{3tanglef}) gives  
\ben
\tau_f(\ket{\psi_{\textsc{w}}})=0=\tau(\ket{\textsc{w}_q}).
\een
This expression, together with (\ref{concufyq}), establishes the identification between the amount of fermionic entanglement present in the indistinguishable-fermionic state (\ref{tipow}), and the amount of (usual) entanglement in the distinguishable-qubit state (\ref{3qu}), that serves as a useful resource for information tasks.

The results also reinforce the correspondence between \textsc{w}-type states in tripartite systems of fermions and qubits, and shows that the latter can be generated from the former via the tunneling plus particle-detection process.

\section{Concluding remarks}

We proposed a protocol based on a tunneling plus a particle-detection process for producing tripartite entangled states out of an initial pure state of three indistinguishable fermions (with a six-dimensional Hilbert space each), exhibiting correlations exclusively due to the antisymmetrization.

The protocol allows to produce fermionic \textsc{w}-type states, which in contrast to the initial Slater determinant, possess a non-zero amount of multipartite fermionic entanglement.
The fermionic entanglement present in the final state is completely transformed into    accessible entanglement among three effective and distinguishable qubits when agents (such as Alice, Bob, and Charlie) have access to particles in their respective wells ($a$, $b$, and $c$).
Under these circumstances, Alice, Bob, and Charlie jointly share a genuinely entangled 3-qubit state that belongs to the \textsc{w}-type family, exhibiting entanglement across all bipartitions (non-vanishing $C_3$), and null 3-tangle.

Notably, the proposed protocol does not allow for the generation of  \textsc{ghz}-type states, i.e., it is unable to generate tripartite states with non-vanishing 3-tangle. 
This shows that the particle detection process, which is ultimately the operation that gives rise to the  multipartite entanglement generation, creates only \emph{certain} classes of multipartite entanglement whenever the preceding unitary evolution $\hat U=\hat U_f\otimes \hat U_f\otimes \hat U_f$ involves operators $\hat U_f$ that factorize as in Eq. (\ref{singleU}) (for more general unitaries $\hat U_f$, more general entangled states may be generated under the  particle detection process).
This  observation goes in line with the fact that the \textsc{w} and the \textsc{ghz}-type states pertain to inequivalents classes of entangled states, not only in systems of distinguishable parties, but also in composites of indistinguishable fermions. 

Our results thus shed light into the characterization of processes that  create a specific type of multipartite entanglement. 
Their implications also  extend to the realm of multipartite entanglement in systems of identical particles, paving the way for further exploration into the role of indistinguishability as a resource for quantum information processing.
 
\begin{acknowledgements} 
M.D.J. and A.P.M. acknowledge funding from Grants
No. PICT 2020-SERIEA-00959 from ANPCyT (Argentina) and No. PIP 11220210100963CO from CONICET (Argentina) and partial support from SeCyT, Universidad Nacional de Córdoba (UNC), Argentina. A.V.H. acknowledges financial support from DGAPA-UNAM through project PAPIIT IN112723. 
\end{acknowledgements}

\appendix
\section{Derivation of equations in Section IIIC}\label{equalcoef}
\subsection{Derivation of equations (\ref{cond})}
From Eqs. (\ref{tunnel}), (\ref{etas}) and (\ref{Tmat}) we get for the coefficients 
$|\eta_i-\eta_j|$ the following:
\begin{widetext}
\begin{subequations}\label{absetas}
\begin{eqnarray}
|\eta_2-\eta_1|^2&=&\sin ^2\theta_2 \sin ^2\theta_4 \left(- \sin 2\theta_2  
\cos \theta_4
\sin \theta_6 
\cos \theta_6 
\cos \theta +\sin ^2\theta_2 \cos ^2\theta_4 \sin ^2\theta_6+\cos ^2\theta_2 \cos ^2\theta_6\right)\nonumber\\
&=&(1-\cos ^2\theta_2) \sin ^2\theta_4 \left[- x +u^2(1-\cos ^2\theta_2) +v^2(1-\sin ^2\theta_2)\right]\\
|\eta_2-\eta_3|^2&=&\cos ^2\theta_2 \sin ^2\theta_4 \left(+2 \sin 2\theta_2  \cos \theta_4 \sin \theta_6 \cos \theta_6 \cos \theta +\cos ^2\theta_2 \cos ^2\theta_4 \sin ^2\theta_6+\sin ^2\theta_2 \cos ^2\theta_6\right),\nonumber\\
&=&\cos ^2\theta_2 \sin ^2\theta_4 \left[ x +u^2\cos ^2\theta_2 +v^2\sin ^2\theta_2\right]\label{coef23}\\
|\eta_1-\eta_3|^2&=&\sin^2\theta_4 \cos^2\theta_4 \sin^2\theta_{6}=u^2\sin^2\theta_4,\label{ape}
\end{eqnarray}
\end{subequations}
with $\theta=2\,(\theta_{3}+\theta_{5})$, $x=uv\sin 2\theta_2   
\cos \theta$, $u=\cos \theta_4 \sin \theta_6$, and $v=\cos \theta_6$.
Notice that consequently the probabilities $|\eta_i-\eta_j|^2$ depend only on four parameters, namely $\theta_2, \theta_4, \theta_6,$ and $\theta$. 
Further, from Eqs. 
 (\ref{absetas}) it follows that for having all three non-vanishing coefficients, i.e., for having a state of the form (\ref{tipow}), it is necessary that 
$\sin\theta_4,\cos\theta_4,\sin\theta_2,\cos\theta_2$, and $\sin\theta_6$ are all non-vanishing terms.

On one hand, from the condition $|\eta_2-\eta_1|^2=|\eta_2-\eta_3|^2$ we get 
\begin{eqnarray}
(1-\cos ^2\theta_2)  \left[- x +u^2(1-\cos ^2\theta_2) +v^2(1-\sin ^2\theta_2)\right]=\cos ^2\theta_2  \left[ x +u^2\cos ^2\theta_2 +v^2\sin ^2\theta_2\right]
\end{eqnarray}
which leads directly to $x=-u^2\cos 2\theta_2$, or rather to 
\ben\label{cond1}
u=-v\tan(2\theta_2)\cos\theta.
\een
Substituting the value of the auxiliary parameters $u,v$, we arrive at Eq. (\ref{condb}). 

On the other hand, and after substituting (\ref{cond1}) into (\ref{coef23}), the condition
$|\eta_2-\eta_3|^2=|\eta_1-\eta_3|^2$ implies
\ben \label{cond2}
\sin^2(2\theta_2)(u^2+v^2)=4u^2.
\een
Combination of (\ref{cond1}) and (\ref{cond2}) leads finally to Eq. (\ref{conda}). 
\end{widetext}

\subsection{Implications of the symmetry condition $|t_{\alpha\beta}|^2=|t_{\beta\alpha}|^2$}

From $|t_{ab}|^2=|t_{ba}|^2$ we get 
\begin{equation}
\sin \theta_4 \left[\cos (2\theta_2)-\cos (2\theta_6)\right]=0,
\end{equation}
whereas for $|t_{ac}|^2=|t_{ca}|^2$ it follows that
\begin{equation}
    \sin \theta_4 \left(\cos ^2\theta_2-\cos ^2\theta_6\right)=0.
\end{equation}
The last condition, $|t_{bc}|^2=|t_{cb}|^2$, gives 
\begin{equation}
    \sin \theta_4 \left(\sin ^2\theta_2-\sin ^2\theta_6\right)=0.
\end{equation}
Since $\sin ^2\theta_4\neq 0$ (as argued below Eqs. (\ref{absetas})), we conclude that the symmetric probability condition $|t_{\alpha\beta}|^2=|t_{\beta\alpha}|^2$ implies
\ben
\theta_2 = \theta_6.
\een
Further, since $\theta_{1}$, $\theta_{7}$, and $\theta_{8}$ are not constrained by the conditions (\ref{cond}), and these latter involve $\theta_3$ and $\theta_5$ only via the sum (\ref{sum}), we can (without loss of generality) set $\theta_{1}=\theta_{7}=\theta_{8}=\theta_{5}=0$

\subsection{Explicit form of the tunneling probability amplitudes}

With the conditions presented in the main text, by using the explicit form of the Gell-Mann matrices (\ref{Tsim}) and by inspection of (\ref{tunnel}), we obtain the following set of equations for the elements $t_{\alpha\beta}$ of $\hat T$:
\begin{eqnarray}\label{eq:tunnelings}
\tilde t_{cc}&=&\cos \theta_4,\nonumber\\
\tilde t_{aa}&=&e^{i \theta_3} \tilde t_{cc} \cos ^2\theta_2 -e^{-i \theta_3} \sin ^2\theta_2,\nonumber \\
\tilde t_{bb}&=&e^{-i \theta_3}  \cos ^2\theta_2 -e^{i \theta_3} \tilde t_{cc}\sin ^2\theta_2,\nonumber \\
\tilde t_{ab}&=&\cos \theta_2 \sin\theta_2(e^{-i\theta_3}+\tilde t_{cc}e^{i\theta_3})=-\tilde t_{ba},\nonumber \\
\tilde t_{ac}&=&e^{i \theta_3} \cos \theta_2 \sin\theta_4=-e^{i \theta_3} \tilde t_{ca},\nonumber \\
\tilde t_{bc}&=&-e^{i \theta_3} \sin \theta_2 \sin \theta_4=e^{i \theta_3} \tilde t_{cb}.
\end{eqnarray}

\begin{thebibliography}{10}

\bibitem{nielsen_chuang_book}
M.~A. Nielsen and I.~L. Chuang, {\em Quantum Computation and Quantum
  Information}.
\newblock Cambridge University Press, 2000.

\bibitem{Jozsa2003}
R.~{Jozsa} and N.~{Linden}, ``{On the role of entanglement in
  quantum-computational speed-up},'' {\em Proceedings of the Royal Society of
  London Series A}, vol.~459, pp.~2011--2032, Aug. 2003.

\bibitem{horodecki_2009}
R.~Horodecki, P.~Horodecki, M.~Horodecki, and K.~Horodecki, ``Quantum
  entanglement,'' {\em Rev. Mod. Phys.}, vol.~81, pp.~865--942, Jun 2009.

\bibitem{zanardi_2000}
P.~Zanardi, C.~Zalka, and L.~Faoro, ``Entangling power of quantum evolutions,''
  {\em Phys. Rev. A}, vol.~62, p.~030301, Aug 2000.

\bibitem{Benatti_2020}
F.~Benatti, R.~Floreanini, F.~Franchini, and U.~Marzolino, ``Entanglement in
  indistinguishable particle systems,'' {\em Physics Reports}, vol.~878,
  pp.~1--27, 2020.
\newblock Entanglement in indistinguishable particle systems.

\bibitem{Dalton_Identical_Bosons_I}
B.~M.~G. B.~J.~Dalton, J.~Goold and M.~D. Reid, ``Quantum entanglement for
  systems of identical bosons: I. general features,'' {\em Physica Scripta},
  vol.~92, no.~2, p.~023004, 2017.

\bibitem{Morris_2020}
B.~Morris, B.~Yadin, M.~Fadel, T.~Zibold, P.~Treutlein, and G.~Adesso,
  ``Entanglement between identical particles is a useful and consistent
  resource,'' {\em Phys. Rev. X}, vol.~10, p.~041012, Oct 2020.

\bibitem{Ghirardi_2002}
G.~Ghirardi, L.~Marinatto, and T.~Weber, ``Entanglement and properties of
  composite quantum systems: A conceptual and mathematical analysis,'' {\em
  Journal of Statistical Physics}, vol.~108, pp.~49--122, 2002.

\bibitem{Majtey2023}
A.~P. {Majtey}, A.~{Vald{\'e}s-Hern{\'a}ndez}, and E.~{Cuestas},
  ``{Indistinguishable entangled fermions: basics and future challenges},''
  {\em Philosophical Transactions of the Royal Society of London Series A},
  vol.~381, p.~20220108, Sept. 2023.

\bibitem{schielmann_2001}
J.~Schliemann, J.~I. Cirac, M.~Ku\ifmmode~\acute{s}\else \'{s}\fi{},
  M.~Lewenstein, and D.~Loss, ``Quantum correlations in two-fermion systems,''
  {\em Phys. Rev. A}, vol.~64, p.~022303, Jul 2001.

\bibitem{Eckert_2002}
K.~Eckert, J.~Schliemann, D.~Bruß, and M.~Lewenstein, ``Quantum correlations
  in systems of indistinguishable particles,'' {\em Annals of Physics},
  vol.~299, no.~1, pp.~88--127, 2002.

\bibitem{plastino_2009_epl}
A.~R. Plastino, D.~Manzano, and J.~S. Dehesa, ``Separability criteria and
  entanglement measures for pure states of n identical fermions,'' {\em {EPL}
  (Europhysics Letters)}, vol.~86, p.~20005, apr 2009.

\bibitem{killoran_2014}
N.~Killoran, M.~Cramer, and M.~B. Plenio, ``Extracting entanglement from
  identical particles,'' {\em Phys. Rev. Lett.}, vol.~112, p.~150501, 2014.

\bibitem{bouvrie_2017_b}
P.~A. Bouvrie, A.~Vald\'es-Hern\'andez, A.~P. Majtey, C.~Zander, and A.~R.
  Plastino, ``Entanglement generation through particle detection in systems of
  identical fermions,'' {\em Ann. Phys.}, vol.~383, p.~401, 2017.

\bibitem{bouvrie_2019}
P.~A. Bouvrie, E.~Cuestas, I.~Roditi, and A.~P. Majtey, ``Entanglement between
  two spatially separated ultracold interacting fermi gases,'' {\em Phys. Rev.
  A}, vol.~99, p.~063601, Jun 2019.

\bibitem{Lange_2018}
K.~Lange, J.~Peise, B.~L{\"u}cke, I.~Kruse, G.~Vitagliano, I.~Apellaniz,
  M.~Kleinmann, G.~T{\'o}th, and C.~Klempt, ``Entanglement between two
  spatially separated atomic modes,'' {\em Science}, vol.~360, no.~6387,
  pp.~416--418, 2018.

\bibitem{Becher_2020}
J.~H. Becher, E.~Sindici, R.~Klemt, S.~Jochim, A.~J. Daley, and P.~M. Preiss,
  ``Measurement of identical particle entanglement and the influence of
  antisymmetrization,'' {\em Phys. Rev. Lett.}, vol.~125, p.~180402, Oct 2020.

\bibitem{Carvalho2004}
A.~R.~R. Carvalho, F.~Mintert, and A.~Buchleitner, ``Decoherence and
  multipartite entanglement,'' {\em Phys. Rev. Lett.}, vol.~93, p.~230501, Dec
  2004.

\bibitem{dur2000}
W.~D\"ur, G.~Vidal, and J.~I. Cirac, ``Three qubits can be entangled in two
  inequivalent ways,'' {\em Phys. Rev. A}, vol.~62, p.~062314, Nov 2000.

\bibitem{verstraete2002}
F.~Verstraete, J.~Dehaene, B.~De~Moor, and H.~Verschelde, ``Four qubits can be
  entangled in nine different ways,'' {\em Phys. Rev. A}, vol.~65, p.~052112,
  Apr 2002.

\bibitem{Coffman2000}
V.~Coffman, J.~Kundu, and K.~Wootters, ``Distributed entanglement,'' {\em
  \href{https://journals.aps.org/pra/abstract/10.1103/PhysRevA.61.052306}{Phys.
  Rev. A}},
  vol.~\href{https://journals.aps.org/pra/abstract/10.1103/PhysRevA.61.052306}{61},
  p.~052306, 2000.

\bibitem{tichy_2011_JPB}
M.~C. {Tichy}, F.~{Mintert}, and A.~{Buchleitner}, ``{Essential entanglement
  for atomic and molecular physics},'' {\em Journal of Physics B Atomic
  Molecular Physics}, vol.~44, p.~192001, Oct. 2011.

\bibitem{majtey_2016}
A.~P. Majtey, P.~A. Bouvrie, A.~Vald\'es-Hern\'andez, and A.~R. Plastino,
  ``Multipartite concurrence for identical-fermion systems,'' {\em Phys. Rev.
  A}, vol.~93, p.~032335, 2016.

\bibitem{debarba2019}
A.~C. Louren\ifmmode~\mbox{\c{c}}\else \c{c}\fi{}o, T.~Debarba, and E.~I.
  Duzzioni, ``Entanglement of indistinguishable particles: A comparative
  study,'' {\em Phys. Rev. A}, vol.~99, p.~012341, Jan 2019.

\bibitem{Levay2008}
P.~L\'evay and P.~Vrana, ``Three fermions with six single-particle states can
  be entangled in two inequivalent ways,'' {\em Phys. Rev. A}, vol.~78,
  p.~022329, Aug 2008.

\bibitem{henderson2009experimental}
K.~Henderson, C.~Ryu, C.~MacCormick, and M.~Boshier, ``Experimental
  demonstration of painting arbitrary and dynamic potentials for bose--einstein
  condensates,'' {\em New Journal of Physics}, vol.~11, no.~4, p.~043030, 2009.

\bibitem{Byrd1998}
M.~{Byrd}, ``{Differential geometry on SU(3) with applications to three state
  systems},'' {\em Journal of Mathematical Physics}, vol.~39, pp.~6125--6136,
  Nov. 1998.

\bibitem{Byrd2000}
M.~{Byrd}, ``{Erratum: ``Differential geometry on SU(3) with applications to
  three state systems'' [J. Math. Phys. 39, 6125-6136 (1998)]},'' {\em Journal
  of Mathematical Physics}, vol.~41, pp.~1026--1030, Feb. 2000.

\bibitem{Cveti2002}
M.~{Cveti{\v{c}}}, G.~W. {Gibbons}, H.~{L{\"u}}, and C.~N. {Pope},
  ``{Cohomogeneity one manifolds of Spin(7) and G$_{2}$ holonomy},'' {\em
  \prd}, vol.~65, p.~106004, May 2002.

\bibitem{Tilma_2002}
T.~Tilma and E.~C.~G. Sudarshan, ``Generalized euler angle parametrization for
  su(n),'' {\em Journal of Physics A: Mathematical and General}, vol.~35,
  p.~10467, nov 2002.

\bibitem{byrd1997geometry}
M.~Byrd, ``The geometry of su(3),'' 1997.

\end{thebibliography}
\end{document}